\documentclass[%
 reprint,
 amsmath,amssymb,
 aps,
]{revtex4-2}

\usepackage{graphicx}
\usepackage{dcolumn}
\usepackage{bm}
\usepackage{multirow}

\providecommand{\keywords}[1]
{
  \small	
  \textbf{\textit{Keywords---}} #1
}

\begin{document}


\title{Estimating the severity of COVID-19 Omicron Variant in USA, India, Brazil, France, Germany and UK}

\author{Tulio Eduardo Rodrigues$^{1,2}$}
\email{tulio@if.usp.br}
\author{Otaviano Helene$^{1}$}
\affiliation{
 $^{1}$Experimental Physics Department, 
Physics Institute, University of São Paulo - Brazil\\
$^{2}$ Safe and Optimized Standards Assessoria Técnica e Diagnósticos LTDA}

\date{\today}

\begin{abstract}
This work evaluates the severity of COVID-19 Omicron variant in terms of the case-fatality-rates (CFR) with respective uncertainty intervals via a simultaneous fitting of confirmed cases and deaths in the USA, India, Brazil, France, Germany and United Kingdom. The CFRs were calculated within the predominance of the Omicron surge combining Monte Carlo simulations and analytical methods based on Gompertz functions. Data on confirmed cases and deaths from all six countries were fitted simultaneously using interactive first order Taylor expansions for the non-linear parameters under the framework of the Least Square Method and assuming that the deaths can be described by a convolution of the confirmed cases with a common gamma function to describe the case to death period. Linear backgrounds both for cases and deaths were included in the fitting to account for the contributions from other strains within the Omicron peaks. The 95\% confidence intervals of the fitted functions and the CFRs were obtained by uncertainty propagation using the full covariance matrix of the best-fit parameters. The fitting included 125 and 113 epidemiological weeks, for cases and deaths, respectively, and 64 parameters fitted simultaneously, resulting in a $\chi ^{2}$ of 176.5 for 174 degrees of freedom ($p = 0.434$). The CFRs with 95\% confidence intervals for USA, India, Brazil, France, Germany and United Kingdom were 0.295 (0.154-0.436)\%, 0.232 (0.134-0.331)\%, 0.49 (0.27-0.71)\%, 0.056 (0.028-0.084)\%, 0.129 (0.074-0.184)\% and 0.168 (0.107-0.229)\%, respectively. The case to death period was satisfactorily described by a common gamma function with mean of $15.71\pm 0.55$ days and coefficient of variation of $0.354\pm 0.070$. The proposed calculations described the dynamics of confirmed cases and deaths during the outbreak of the Omicron variant of COVID-19, providing accurate and reliable information about the respective CFRs and the case to death period, the latter being consistent with previous estimates for the symptom onset to death made in the early stages of the pandemic. The CFRs thus obtained are considerably lower than previous measurements available in the literature, suggesting that the latter may have been overestimated, as the probability of deaths from other strains of the virus under the generally prominent Omicron peak, here accounted for in terms of linear backgrounds, was not considered. The method can also be applied in other pandemics, contributing to the understanding of their dynamics and severity, and helping decision makers in the implementation of effective policies to face similar diseases around the world.  

\end{abstract}


\maketitle


\section{Introduction}

The advent of the COVID-19 pandemic has generated an unprecedented scenario for human beings, representing a major epidemiological risk worldwide \cite{wu2020new,ciotti2020covid}. There was a great commitment from the international scientific community on several research fronts, involving the search for drugs \cite{sanders2020pharmacologic}, the importance of implementing non-pharmacological measures \cite{flaxman2020report, perra2021non}, the development of safe and effective vaccines \cite{ndwandwe2021covid}, the role and predictive capacity of mathematical models \cite{meehan2020modelling, vespignani2020modelling}, among many others.\\
The first confirmed case of Sars-Cov-2 Omicron Variant (B.1.1.529) was reported in South Africa on November 21, 2022, generating a global public health concern \cite{vaughan2021omicron} given the uncertainties with its severity, transmission rate and ability to escape from immunization, either acquired from vaccines or previous infections.\\
Specifically with regard to mathematical developments, it was found that even the most sophisticated epidemiological models were not able to make very accurate predictions over several weeks, especially with regard to estimates of uncertainty intervals \cite{eker2020validity}. The great complexity of the virus propagation dynamics and the lack of knowledge of some parameters make it very difficult to apply epidemiological models to quantitatively describe the evolution of the pandemic. In fact, the number of unreported cases and the high proportion of asymptomatic infections for COVID-19 generate huge uncertainties to determine the actual size of the pandemic along time, which is a crucial part of any epidemiological model.\\
A survey aimed at accessing real-time community transmission [REal-time Assessment of Community Transmission-1 (REACT-1)] in the United Kingdom used RT-PCR tests on a random sample of 94,950 individuals and found a prevalence of 2.88\% for COVID-19 between February 8 and March 1, 2022 \cite{chadeau2022omicron}. Considering a population of 67 million and confirmed cases of COVID-19, Sigal, Milo, and Jassat estimated a total of 73\% unreported infections in the UK \cite{sigal2022estimating}, making difficult the use of mathematical approaches based on the numbers of infected and susceptible people. Alternatively, phenomenological and simpler models can be quite useful to describe a pandemic, provided that they use adequate functionals with parameters consistently fitted to the available data.\\
The severity of COVID-19 during the early stages of the pandemic, expressed in terms of the infection-fatality-rates (IFR), was reasonably well determined analysing the Wuhan data \cite{verity2020estimates}. More specifically, the age-dependent IFR obtained in this study provided suitable constraints to infer the dynamics of true infections through the dynamics of deaths \cite{flaxman2020report,rodrigues2020monte}. However, for Omicron variant there are only few estimates of the infection-fatality-rates with very small figures \cite{chen2022estimation,erikstrup2022seroprevalence}. For instance, Chen \textit{et al.} \cite{chen2022estimation} reported IFR values of 0.01\%, 0.13\% and 1.99\% for ages 40 to 59, 60 to 79 and above 80 years, respectively, for the Omicron BA.2 sublineage outbreak in Shanghai from February-June 2022; evidencing, the much higher severity of the disease for elderly people and figures consistent with zero for ages below 40. Moreover, the IFR results increase one order of magnitude comparing the ages between 60 to 79 years with ages above 80 years, making it very difficult to extrapolate these results for other countries, as performed elsewhere during the early stages of the pandemic \cite{rodrigues2020monte}. Moreover, Erikstrup \textit{et al.} \cite{erikstrup2022seroprevalence} reported an overall IFR of 0.0062\% in a seroprevalence study based on a sample of healthy Danish blood donors between 17 to 72 years old, which, despite being a quite relevant measurement, does not allow any extrapolation to the general population.\\
So, in order to overcome the difficulties to estimate true infections for Omicron, this work proposes a novel approach - based on Monte Carlo and analytical methods - to simultaneously describe the evolution of confirmed cases and deaths using Gompertz functions, thus providing plausible estimates for the severity of this variant in few countries. The calculation includes data from USA, India, Brazil, France, Germany and UK within the Omicron variant predominance and considers a common gamma distribution for all six countries to account for the case to death period, with average value, coefficient of variation (CV) and their respective covariance matrix being directly obtained from the fitting. The method is particularly useful for the calculation of the uncertainties of the best-fit parameters and the corresponding confidence intervals of the fitted functions, which, in turn, are of paramount relevance to help decision-makers and governments to face a pandemic.
\section{Methods}
\subsection{The role of Gompertz functions to describe confirmed cases and deaths of COVID-19}
Following the same steps described elsewhere \cite{rodrigues2020monte}, the total number of infected people at time \textit{t} can be satisfactorily described by a Gompertz function, such that:
\[
I(t)=N_{I}e^{-e^{-\lambda (t-t_{0})}},
\]
where $N_{I}$ represents the asymptotic total number of infections, $\lambda$ the growth rate and $t_{0}$ the inflection point, which corresponds to the peak of new infections per time unit, which is given by the derivative of \textit{I(t)} as follows: 
\begin{equation}
n_{I}(t)=\dfrac{d}{dt}I(t) = N_{I}\lambda e^{-e^{-\lambda (t-t_{0})}}e^{-\lambda (t-t_{0})}.
\end{equation}

Assuming that the number of new confirmed cases also follows the derivative of a Gompertz function, and considering the possibility of linear backgrounds in the data, one may find the following expression for the new cases at time \textit{t}:
\begin{equation}
\label{Eq.2}
\begin{aligned}
& F_{c}(t)= N_{c}\lambda e^{-e^{-\lambda (t-t_{0})}}e^{-\lambda (t-t_{0})} + C_{c} + S_{c}t \\
&\quad = N_{c}f_{c}(t,t_{0}, \lambda)+ C_{c} + S_{c}t,
\end{aligned}
\end{equation}
where $ N_{c}$ represents the asymptotic total number of confirmed cases, $ C_{c}$ a positive constant number of new cases, $S_{c}$ a slope for new cases along time, which can be positive or negative, with $f_{c}(t,t_{0}, \lambda)=\lambda e^{-e^{-\lambda (t-t_{0})}}e^{-\lambda (t-t_{0})}$ being the normalized derivative of the Gompertz function for confirmed cases.\\
The hypothesis that the confirmed cases curves can be satisfactorily described by the derivative of a Gompertz function holds reasonably well as far as the reporting methods and policies do not vary considerably within the desired time interval. Such hypothesis seems plausible given the relatively small time window for the peaks dominated by the Omicron variant (see the Discussion section). Furthermore, the linear background in Eq. \ref{Eq.2} can be interpreted as the new cases related with other variants of the virus, under the predominance of the Omicron peak. As discussed later, the inclusion of linear backgrounds both for confirmed cases and deaths remarkably improves the overall fitting.\\
As similarly proposed elsewhere \cite{flaxman2020report,rodrigues2020monte},  the number of deaths along time can be described by the convolution of the normalized distribution of confirmed cases with a gamma function $G(t,\alpha,\beta)$ to account for the case to death period, such that:
\begin{equation}
\begin{aligned}
\label{Eq.3}
&F_{d}(t)={N_{d}\int_{0}^{t}f_{c}(\tau,t_{0}, \lambda)G(t-\tau,\alpha,\beta)}d\tau+ C_{d} + S_{d}t\\
&\quad = N_{d}f_{d}(t,t_{0},\lambda, \alpha, \beta)+C_{d} + S_{d}t,
\end{aligned}
\end{equation}
where $ F_{d}(t)$ is the number of new deaths at time \textit{t}, $N_{d}$ the asymptotic total number of deaths, $C_{d} $ a positive constant and $ S_{d}$ a slope for deaths, with:
\begin{equation}
\begin{aligned}
G(t,\alpha,\beta)= \frac{\beta^{\alpha}t^{\alpha-1}e^{-\beta t}}{\Gamma(\alpha)},
\end{aligned}
\end{equation}
and
\begin{equation}
\begin{aligned}
f_{d}(t,t_{0}, \lambda, \alpha, \beta)={\int_{0}^{t}f_{c}(\tau,t_{0}, \lambda)G(t-\tau,\alpha,\beta)}d\tau.
\end{aligned}
\end{equation}

The mean value and coefficient of variation of the gamma function are given by $ \mu = \frac{\alpha}{\beta}$ and $\textsc{CV} = \frac{1}{\sqrt{\alpha}}$, whereas $ f_{d}(t,t_{0},\lambda, \alpha, \beta)$ represents the normalized function for the deaths peak, which carries all the shape information from the respective confirmed cases peak. Once again, linear backgrounds were included in Eq. \ref{Eq.3} to account for the likelihood of deaths, under the usually prominent peak, not necessarily related with Omicron infections. Such method allows the disentanglement of the Omicron peak, mathematically connected with the confirmed cases peak through the time-constraint given by $G(t,\alpha,\beta)$, from the actual deaths curves. Under this construction, the case-fatality-rate (CFR) for Omicron can be calculated simply by the ratio between the normalization of deaths and confirmed cases, such that:
\begin{equation}
\begin{aligned}
\textrm{CFR} = \dfrac{N_{c}}{N_{d}}.
\end{aligned}
\end{equation}
As described in more detail later, the uncertainty in CFR depends on the full covariance matrix of all the parameters of the fitting, including the normalization factors $N_{c}$ and $N_{d}$, the linear backgrounds both for cases and deaths and the non-linear parameters of the Gompertz ($\lambda$ and $t_{0}$) and gamma ($\alpha$ and $\beta$) functions.\\
In some countries that presented more than one peak for confirmed cases within the Omicron surge, additional Gompertz functions were included to achieve a successful fitting (see the Results and Discussion sections), such that:
\begin{equation}
\begin{aligned}
& F_{c}(t) = N_{c1}f_{c}(t,t_{01},\lambda_{1})+ N_{c2}f_{c}(t,t_{02},\lambda_{2}) + ...\\
&\quad + C_{c} + S_{c}t,
\end{aligned}
\end{equation}
and
\begin{equation}
\begin{aligned}
& F_{d}(t)=N_{d1}f_{d}(t,t_{01},\lambda_{1},\alpha,\beta)+\\
&\quad N_{d2}f_{d}(t,t_{02},\lambda_{2},\alpha,\beta)+...+C_{d} + S_{d}t,
\end{aligned}
\end{equation}
where $ N_{ci}$ and  $ N_{di} $ $ (i=1,2,3...) $ refer, respectively, to the normalization factors for all the peaks for confirmed cases and deaths with $ t_{0i} $ and $ \lambda_{i} $.
\subsection{Data fitting}

The fitting includes the weekly-averaged daily confirmed cases and deaths for the six countries with the highest number of confirmed COVID-19 cases by April 25, 2022, namely, the United States ($\sim $ 80 M), India ($\sim $ 43 M), Brazil ($\sim $ 30 M), France ($\sim $ 27 M), Germany ($\sim $ 24 M) and United Kingdom ($\sim $ 22 M), according with the World Health Organization webpage (https://covid19.who.int/table). The time intervals were chosen to capture the early stages of Omicron infections in each country up to the last epidemiological week ending on the cut-off date of April 25, 2022. In that sense, the first epidemiological weeks for confirmed cases were chosen to start on November 23, 2011 for the USA; November 30, 2021 for Brazil, France and UK and; December 07, 2021 for India and Germany. The first epidemiological weeks for deaths were chosen 14 days after the weeks for cases for each country.\\
As described in a previous calculation \cite{rodrigues2020monte}, the use of weekly-averaged daily data avoids the trivial seasonal oscillations that appear due to the usually lower effectiveness of reports during the weekends. This approach also provides reasonable estimates for the uncertainties of the weekly-averages, which were assumed as the corresponding standard errors.\\
The fitting procedure was performed in a two-step process. Firstly, the confirmed cases data were fit using a Monte Carlo algorithm \cite{rodrigues2020monte} to sort plausible candidates for the parameters $ N_{c} $, $ t_{0} $, $ \lambda $, $ C_{c}$ and $ S_{c}$ of Eq. \ref{Eq.2} until reaching the lowest $\chi^2$ for each country $s$, defined as:
\begin{equation}
\chi^{2}_{s}=\sum_{ics=1}^{nc_{s}}\frac{[ (\tilde{F_{c}})_{ics} - (y_{c})_{ics}]^{2}}{[(\sigma y_{c})_{ics}]^{2}},
\end{equation}
where $(\tilde{F_{c}})_{ics}$ is the trial function for cases calculated at the mean days of each epidemiological week ($t_{ics}$) up to the last week for cases $nc_{s}$, with $(y_{c})_{ics}$ representing the weekly-averaged cases per day with their corresponding standard errors $(\sigma y_{c})_{ics}$, all from country $s$. \\
The second step of the fitting includes both the confirmed cases and deaths data and assumes that the non-linear parameters $t_{0}$ and $\lambda$ obtained during the first step are sufficiently close to the best-fit parameters $\tilde{t_{0}}$ and $\tilde{\lambda}$. Also, during this step, the initial parameters of the gamma function were chosen at $\alpha=4.938$ and $\beta=0.277$ $d^{-1}$, thus, yielding a mean of 17.8 days and CV of 0.45, as previously found for the symptom onset to death period in the analysis of Wuhan data \cite{verity2020estimates}. In fact, it is expected that the case to death period will be similar but shorter than the symptom onset to death (see the Discussion section). So, under this approximation one can write: $\tilde{t_{0}}\approx t_{0} +\delta t_{0}$, $\tilde{\lambda}\approx \lambda +\delta \lambda$, $\tilde{\alpha}\approx \alpha +\delta \alpha$ and $\tilde{\beta}\approx \beta +\delta \beta$, with $\delta t_{0}\ll t_{0}$, $\delta \lambda \ll \lambda$, $\delta \alpha \ll \alpha$ and $\delta \beta \ll \beta$. Consequently, the following expressions hold:
\begin{equation}
\label{Eq.8}
\begin{aligned}
f_{c}(t,\tilde{t_{0}}, \tilde{\lambda}) \approx f_{c}(t,t_{0}+\delta t_{0}, \lambda + \delta \lambda)
\end{aligned}
\end{equation}
and

\begin{equation}
\label{Eq.9}
\begin{split}
& f_{d}(t,\tilde{t_{0}}, \tilde{\lambda}, \tilde{\alpha}, \tilde{\beta}) \approx \\
 & \quad {\int_{0}^{t}f_{c}(\tau,t_{0}+\delta t_{0}, \lambda + \delta \lambda)G(t-\tau,\alpha + \delta \alpha,\beta +\delta \beta)}d\tau. 
\end{split}
\end{equation}

Making Taylor expansions around $t_{0}$, $\lambda$, $\alpha$ and $\beta$ and collecting only first order terms in $\delta t_{0}$, $\delta \lambda$, $\delta \alpha$ and $\delta \beta$ one can rewrite Eqs. \ref{Eq.8} and \ref{Eq.9}, after some manipulations, as follows:

\begin{equation}
\label{Eq.10}
\begin{aligned}
& f_{c}(t,\tilde{t}_{0}, \tilde{\lambda}) \approx  f_{c}^{0}(t) + f_{c}^{0}(t)\lambda h^{0}(t)\delta t_{0}\\
& \quad + f_{c}^{0}(t)[\frac{1}{\lambda}- (t-t_{0})h^{0}(t)] \delta \lambda,
\end{aligned}
\end{equation}

and

\begin{equation}
\label{Eq.11}
\begin{aligned}
& f_{d}(t,\tilde{t}_{0},\tilde{\lambda}, \tilde{\alpha}, \tilde{\beta})  \approx  \int_{0}^{t}{f_{c}^{0}(\tau)G^{0}(t-\tau) d\tau}\\
& \quad + [\int_{0}^{t}{f_{c}^{0}(\tau)\lambda h^{0}(\tau)G^{0}(t-\tau)d\tau }] \delta t_{0}\\
& \quad +\left\lbrace  \int_{0}^{t}{f_{c}^{0}(\tau)[\frac{1}{\lambda}-(\tau-t_{0})h^{0}(\tau)] G^{0}(t-\tau) d\tau} \right\rbrace \delta \lambda\\
& \quad + \left\lbrace  \int_{0}^{t}{ f_{c}^{0}(\tau)[ln \beta -ln \alpha +ln(t-\tau)]G^{0}(t-\tau) d\tau}\right\rbrace \delta\alpha\\
& \quad +[\int_{0}^{t}{ f_{c}^{0}(\tau)\left( \frac{\alpha}{\beta}-t+\tau\right)G^{0}(t-\tau)d\tau}] \delta\beta,
\end{aligned}
\end{equation}
where $f_{c}^{0}(t)\equiv f_{c}(t, t_{0}, \lambda)$, $h^{0}(t) \equiv 1-e^{-\lambda(t-t_{0})}$ and $G^{0}(t-\tau) \equiv G(t-\tau, \alpha, \beta)$. 

Inserting the expressions found for $f_{c}(t,\tilde{t}_{0}, \tilde{\lambda})$ and $f_{d}(t,\tilde{t}_{0},\tilde{\lambda}, \tilde{\alpha}, \tilde{\beta})$ into Eqs. \ref{Eq.2} and \ref{Eq.3}, respectively, one can write the time evolution of confirmed cases and deaths in terms of known quantities $t_{0}$, $\lambda$, $\alpha$ and $\beta$ and the linear parameters $N_{c}$, $C_{c}$, $S_{c}$, $N_{d}$, $C_{d}$, $S_{d}$, $\delta t_{0}$, $\delta \lambda$, $\delta \alpha$ and $\delta \beta$.\\
In this regard, linear relationships between the functions to be fit and its corresponding parameters were achieved both for confirmed cases and deaths, thus, ensuring the optimal properties of the Least Square Method (LSM) \cite{helene2016useful} related with unbiased estimates with minimal variances. So, during the second step, a global new $\chi^2$ can be calculated using matrix formalism and taking into account the data of confirmed cases and deaths of all six countries simultaneously, as follows:
\begin{equation}
\label{Eq.14}
\begin{aligned}
\chi^2=(Y-XB)^{\top}V^{-1}(Y-XB),
\end{aligned}
\end{equation}
where $X$ is the design matrix with $X_{ics,j} =\frac{\partial F_{c}(t_{ics})}{\partial P_{j}}$ for cases and $X_{ids,j} =\frac{\partial F_{d}(t_{ids})}{\partial P_{j}}$ for deaths, $ P_{j} $ the $ j- $parameter to be fitted, $Y$ the data vector with $ Y_{ics} =(y_{c})_{ics} $ for cases and $ Y_{ids} =(y_{d})_{ids} $ for deaths and $ V $ the diagonal variance matrix with $ V_{ics,ics}=(\sigma y_{c})_{ics}^{2} $ for cases and $ V_{ids,ids} = (\sigma y_{d})_{ids}^{2} $ for deaths. In some datasets, additional uncertainties with magnitudes of 1 to 2\% of the corresponding Gompertz peak heights (either for cases, deaths or both) were added quadratically with $\sigma y_{c} $ or $ \sigma y_{d} $ in order to achieve a successful overall fitting (see the Discussion section). The indexes $ ics $ and $ ids $ range from the first up to the last epidemiological week for cases and deaths, respectively, for each country $ s $. The vector $ B $ represents the fitting parameters and can be calculated as $ B=VbX^{\top}V^{-1}Y $, with $Vb=(X^{\top}V^{-1}X)^{-1}  $ being its corresponding covariance matrix.\\
During the second step of the fitting, the non-linear parameters are replaced by $\tilde{t_{0}}=t_{0}+\tilde{\delta t_{0}}$, $\tilde{\lambda}=\lambda+\tilde{\delta\lambda}$, $ \tilde{\alpha}=\alpha+\tilde{\delta\alpha} $ and $ \tilde{\beta}=\beta+\tilde{\delta\beta} $ in the design matrix and the fitting continues iteratively until $ \tilde{\delta t_{0}}\approx\tilde{\delta\lambda}\approx\tilde{\delta\alpha}\approx\tilde{\delta\beta}\approx0 $, thus, resulting in the best-fit parameters to describe cases and deaths of all the countries simultaneously.  

\section{Results}
The results of the initial fitted parameters during the Monte Carlo sampling are presented in Table \ref{table1} and were obtained by mapping the lowest $ \chi ^{2}$ for each country. The parameters were varied uniformly $ 10^{6} $ times around plausible guessing intervals and the best combination, which matches the lowest $\chi^{2}$ for each country, was selected. For USA, India and Brazil single peaks were included to fit the confirmed cases data, whereas for France, Germany and UK multiple peaks were included. All the parameters are shown in Table \ref{table1} with three significant algharisms.\\
The second fitting, herein denoted global fitting, was performed using the LSM with matrix formalism (Eq.\ref{Eq.14}) and the initial fitted parameters $t_{0}$ and $\lambda$ for each Gompertz peak of each country depicted in Table \ref{table1}; also assuming the initial values of the gamma function $\alpha=4.938$ and $\beta=0.277$ $d^{-1}$. The results of the global fitting are shown in Table \ref{table2}, and include a total of 64 parameters simultaneously fitted to describe the weekly-averaged daily confirmed cases and deaths for all six countries.\\
Figures 1 and 2 present, respectively, the results of the global fitting for confirmed cases and deaths (solid blue lines) in comparison with the available data (black squares with error bars) for all six countries, where the dashed-dotted red and green lines represent, respectively, the upper and lower limits of the fitted functions considering a 95\% confidence interval (CI). These limits were calculated by uncertainty propagation, taking into account the full covariance matrix of the fitted parameters, namely $N_{c}$, $\lambda_{i}$, $t_{0i}$, $C_{c}$ and $S_{c}$ for cases and $N_{d}$, $\lambda_{i}$, $t_{0i}$, $C_{d}$, $S_{d}$, $\alpha$ and $\beta$ for deaths, where the index $i$ accounts for the fits that include more than one Gompertz function for cases (France, Germany and UK) and deaths (France and UK) (see the Discussion section).\\
As clearly presented in Figures 1 and 2 and Table \ref{table2}, all the data of cases and deaths are satisfactorily fitted by the proposed model ($\chi^{2}=176.5$ and $D.O.F.=174$). Under this approach, one can estimate with reasonable accuracy the mean value and coefficient of variation of a common gamma function that describes the case to death period simultaneously for all six countries within the Omicron surge. The result of this estimate is presented by the solid blue line of Figure 3, with the dashed-dotted red and green lines representing, respectively, the upper and lower limits of the gamma function considering a 95\% CI.\\
For the countries with two different CFR values (France and UK) we have combined both results for each country in order to achieve a single one. This procedure took into account the covariance matrix $V_{abcd}$ defined as:

\begin{equation}
\label{Eq.15}
\begin{aligned}
V_{abcd} = \begin{bmatrix}
\sigma_{a}^{2} & cov(a,b) & cov(a,c) & cov(a,d)\\
cov(b,a)& \sigma_{b}^{2} & cov(b,c)&
cov(b,d)\\
cov(c,a) & cov(c,b) & \sigma_{c}^{2} &
cov(c,d)\\
cov(d,a) & cov(d,b) & cov(d,c) &
\sigma_{d}^{2}
\end{bmatrix},
\end{aligned}
\end{equation}
where $a\equiv N_{d1}$, $b \equiv N_{c1}$, $c \equiv N_{d2} $ and $d \equiv N_{c2}$, such that $\textrm{CFR}_{1} =a/b$ and $\textrm{CFR}_{2} = c/d$ either for France or UK. The non-diagonal terms of $V_{abcd}$ are the covariances between the fitted parameters. So, the variance matrix of $\textrm{CFR}_{1}$ and $\textrm{CFR}_{2}$ for each country can be calculated as $V_{12}=DV_{abcd}D ^{\top}$, where $D$ is written as:

\begin{equation}
\label{Eq.16}
\begin{aligned}
D = \begin{bmatrix}
\dfrac{\partial \textrm{CFR}_{1} }{\partial a} & \dfrac{\partial \textrm{CFR}_{1} }{\partial b} & \dfrac{\partial \textrm{CFR}_{1} }{\partial c} & \dfrac{\partial \textrm{CFR}_{1} }{\partial d}\\[1.2em]
\dfrac{\partial \textrm{CFR}_{2} }{\partial a} & \dfrac{\partial \textrm{CFR}_{2} }{\partial b} & \dfrac{\partial \textrm{CFR}_{2} }{\partial c} & \dfrac{\partial \textrm{CFR}_{2} }{\partial d}\\
\end{bmatrix}.
\end{aligned}
\end{equation} 

Then, the final CFR values can be calculated as: $\textrm{CFR}=V_{F}X_{F}^{\top} V_{12}^{-1}Y_{F}$, where $V_{F} = (X_{F}^{\top} V_{12}^{-1}X_{F})^{-1}$, $X_{F}=\big(\begin{smallmatrix} 1\\ 1 \end{smallmatrix}\big)$ and $Y_{F}=\big(\begin{smallmatrix} \textrm{CFR}_{1}\\ \textrm{CFR}_{2} \end{smallmatrix}\big)$; thus, resulting $0.056 \pm 0.014$ \% for France and $0.168 \pm 0.031$ \% for UK. 

Figure 4 summarizes the case-fatality-rates (in \%) with their respective 95\% confidence intervals obtained in this model calculation for all six countries within the Omicron predominance.

\section{Discussions}
As observed in Figures 1 and 2, the method herein proposed effectively describes data on cases and deaths via a simultaneous fitting procedure, shedding light on the severity of COVID-19 Omicron variant in USA, India, Brazil, France, Germany and UK. The approach is also advantageous to extract with good accuracy the mean value and respective coefficient of variation of a gamma function capable of describing and connecting cases and deaths data of the six countries simultaneously, which represents an important pandemic parameter not yet obtained for the Omicron variant. Moreover, the Taylor expansions in the non-linear parameters of the Gompertz ($\lambda$ and $t_{0}$) and gamma ($\alpha$ and $\beta$) functions propitiated a framework of linear parameters, ensuring the optimal properties of the Least Square Method \cite{helene2016useful} related with unbiased estimates with minimal variances.\\
As pointed out previously, the confirmed cases can be satisfactorily described by the derivative of a Gompertz like function provided that relevant characteristics of the pandemic, such as: vaccination coverage, reporting methods, non-pharmacological interventions, predominant strain, to name a few, do not change considerably during the period of the fitting. Indeed, the peaks of confirmed cases during Omicron prevalence typically last between two to four months, depending on the country (see Fig. 1), which are short periods for structural changes to occur in public health policies, thus reinforcing the viability and applicability of the proposed model.\\
Regarding the fitting procedure, additional uncertainties proportional to the respective peak heights were arbitrarily added to fit the confirmed cases from USA (1\%), India (2\%) and Brazil (1\%) and the deaths from India (2\%). Despite having a noticeable effect in lowering the overall $\chi^2 $ of the fit, these additional uncertainties did not significantly affect the main parameters of the pandemic waves, such as the peak days, the peak heights and the growth rates.\\
The fits of confirmed cases from France and Germany took into account two well defined peaks, but the deaths data from Germany do not show two distinguished peaks as one would have expected. In fact, the fitting with two peaks for deaths from Germany was discarded, since it generated one peak with magnitude consistent with zero [$(1 \pm10)\times 10^{2}$] and a marginal reduction of 0.02 in the $\chi^{2}$. These difference between the structures of the deaths data from France and Germany could be explained by the smaller time difference between the confirmed cases peaks for Germany (42 days), when compared with France (70 days). For UK, the confirmed cases present two well defined peaks and one small shoulder between them, making it necessary to include a third peak (not consistent with zero) in order to achieve a successful fitting. In this case, only two peaks for deaths were included, since the third one was consistent with zero [$ (-15\pm22)\times10$] with a small reduction of 0.45 in the $\chi^{2}$.\\   
The relative uncertainties of the best-fit parameters (see Table \ref{table2}) range from 5 to 19\% for $N_{c}$, 0.6 to 25\% for $t_{0}$, 5 to 18\% for $\lambda$, 4.5 to 13.3\% for $C_{c}$, 17 to 63\% for $S_{c}$,19 to 62\% for $N_{d}$, 7 to 14\% for $C_{d}$, 9 to 52\% for $S_{d}$ and 18 to 25\% for CFR. Moreover, the mean value of the gamma function has a small relative uncertainty of 3.5\%, when compared with the huge uncertainties of $\alpha $ (39\%) and $\beta $ (41\%), evidencing the need to account for the full covariance matrix when propagating the uncertainties, given the strong correlation between $\alpha$ and $\beta$ [$ \rho_{\alpha,\beta}=cov(\alpha,\beta)/\sigma_{\alpha}\sigma_{\beta}=0.997$]. A similar situation also occurs for the CFRs, which depend on the ratios $N_{d}/N{c}$ for each country. While $N_{d}$ and $N_{c}$ have relative uncertainties ranging from 5 up to 62\%, their ratios have much smaller uncertainties due to the large correlations between the parameters.\\
The fitted gamma distribution shown in Fig. 3 describes the case to death period and represents a reasonable proxy for the symptom onset to death. The average value for $G(t,\alpha, \beta)$ found in this study [15.7 (95\% CI: 13.5 - 17.9 days] is two days shorter, but statistically consistent with previous estimates for the symptom onset to death made in the early stages of the pandemic \cite{verity2020estimates} [17.8 (95\% CI: 16.9 - 19.2) days]. In this regard, the confirmed cases in the dataset of all six countries are being reported with an average delay of $\sim 2$ days from symptom onset, which seems plausible. Moreover, individuals more susceptible for deaths due to the Omicron strain would also present similar profiles of the symptom onset to death period when compared with the early stages of the pandemic.\\
Regarding the severity of the Omicron variant, expressed in terms of the CFRs, the figures found in this study are considerably lower than previous estimates. Specifically for the USA, previous studies have reported CFRs between $\sim $ 0.3 and 1.7\% for the Omicron \cite{sigal2022estimating,cuadros2022association, barletta2022influence}. For instance, Sigal \textit{et al.} \cite{sigal2022estimating} reported a value of $\sim 0.6\%$, which is a factor 2 higher than the one obtained in this study ($0.295\pm0.071$)\%. Similarly, Cuadros \textit{et al.} \cite{cuadros2022association} and Barletta \cite{barletta2022influence} reported figures between $\sim $ 0.3 to 0.7\% and $ \sim $ 0.5 to 1.7 \%, respectively, still well above the value obtained in this work. Furthermore, considering the UK data, Emani \textit{et al}. \cite{emani2022increasing} also found higher CFR results during the first 12 (0.19\% from Dec 6, 2021 to Feb 27, 2022) and the last 9 weeks (0.41\% from Feb 28 to May 1, 2022) of the Omicron outbreak, when compared with the result obtained in the present analysis ($0.168\pm0.031$)\%.\\
Moreover, an analysis based on data from 50 countries during the Omicron outbreak estimated an overall CFR of 0.304 \% (interquartile range: 0.187-0.748\%), which is consistent with the values found in this work for USA, India and Brazil, but still a factor $\sim2$ to $\sim 5$ higher than the results found for France, Germany and the United Kingdom. It is worth mentioning, however, that comparisons between CFR results from different countries with any average value, as the one reported by Wang \textit{et al}. \cite{wang2022differences}, should be done with caution due to the significant differences between countries regarding age distribution, vaccination coverage, hospital facilities, reporting methods, non-pharmacological policies, among others.\\
In summary, the usually lower CFR values found in the present study suggest that the severity of the Omicron variant may have been overestimated in previous calculations, as the backgrounds were not taken into account in the analyses. Consequently, significant contributions to the number of deaths due to other strains of the virus may not have been properly disentangled under the prominent peaks still largely dominated by the Omicron strain, increasing the case fatality rates. Moreover, the inclusion of the linear backgrounds both for cases and deaths avoids the likelihood of artefacts when calculating the CFR. As recently shown in a meta-analysis that included 24 countries \cite{kim2023case}, the CFR for Omicron presented relevant differences between different stages of the pandemic waves. For instance, the CFR obtained during the decreasing phase of the pandemic waves was almost a factor 2 (OR: 1.962, 95\% CI 1.607 - 2.397) higher than the value found during the increasing phase of the Omicron peaks, indicating a relevant CFR dependence on the time interval of the analyses. A similar finding also appeared in the analysis proposed by Emani et al. \cite{emani2022increasing}, where the CFR during the last 9 weeks of Omicron predominance was a factor two higher than the result found in the first 12 weeks. A plausible explanation for these temporal variations in the CFR values may be the likelihood of unaccounted deaths related to infections by other more severe strains that occurred during Omicron predominance. In that sense, the linear backgrounds included in the present analysis represent a consistent phenomenological approach to account for deaths (and cases) from other strains, yielding constant CFR values for each country during the Omicron outbreak.\\
\section{Conclusions}  
This study provides a framework to calculate the case-fatality-rates of COVID-19 Omicron variant and their respective confidence intervals using Monte Carlo simulations and analytical methods based on Gompertz functions. The approach included data on confirmed cases and deaths from the USA, India, Brazil, France, Germany and United Kingdom that were fitted simultaneously using the Least square Method and adopting a common gamma function to describe the case to death period and first order Taylor expansions of the non-linear parameters.\\
The CFRs obtained here are considerably lower than previous estimates, likely due to the inclusion of linear backgrounds both for cases and deaths to account for contributions from other strains under the generally dominant Omicron peak, thus suggesting that Omicron severity may have been overestimated until now.\\
The case to death period was determined for the first time for Omicron and has a mean [15.7 (95\% CI: 13.5 - 17.9 days] consistent with previous estimates for the symptom onset to death made in the early stages of the pandemic [17.8 (95\% CI: 16.9 - 19.2) days]\\
In summary, this new approach allowed a comprehensive and accurate description of the dynamics of confirmed cases and deaths of COVID-19 within the Omicron predominance that can be extended to other pandemics and scenarios, representing a powerful tool to address public health issues and guide decision makers worldwide.

\bibliography{bmc_article}

\begin{thebibliography}{22}%
\makeatletter
\providecommand \@ifxundefined [1]{%
 \@ifx{#1\undefined}
}%
\providecommand \@ifnum [1]{%
 \ifnum #1\expandafter \@firstoftwo
 \else \expandafter \@secondoftwo
 \fi
}%
\providecommand \@ifx [1]{%
 \ifx #1\expandafter \@firstoftwo
 \else \expandafter \@secondoftwo
 \fi
}%
\providecommand \natexlab [1]{#1}%
\providecommand \enquote  [1]{``#1''}%
\providecommand \bibnamefont  [1]{#1}%
\providecommand \bibfnamefont [1]{#1}%
\providecommand \citenamefont [1]{#1}%
\providecommand \href@noop [0]{\@secondoftwo}%
\providecommand \href [0]{\begingroup \@sanitize@url \@href}%
\providecommand \@href[1]{\@@startlink{#1}\@@href}%
\providecommand \@@href[1]{\endgroup#1\@@endlink}%
\providecommand \@sanitize@url [0]{\catcode `\\12\catcode `\$12\catcode
  `\&12\catcode `\#12\catcode `\^12\catcode `\_12\catcode `\%12\relax}%
\providecommand \@@startlink[1]{}%
\providecommand \@@endlink[0]{}%
\providecommand \url  [0]{\begingroup\@sanitize@url \@url }%
\providecommand \@url [1]{\endgroup\@href {#1}{\urlprefix }}%
\providecommand \urlprefix  [0]{URL }%
\providecommand \Eprint [0]{\href }%
\providecommand \doibase [0]{https://doi.org/}%
\providecommand \selectlanguage [0]{\@gobble}%
\providecommand \bibinfo  [0]{\@secondoftwo}%
\providecommand \bibfield  [0]{\@secondoftwo}%
\providecommand \translation [1]{[#1]}%
\providecommand \BibitemOpen [0]{}%
\providecommand \bibitemStop [0]{}%
\providecommand \bibitemNoStop [0]{.\EOS\space}%
\providecommand \EOS [0]{\spacefactor3000\relax}%
\providecommand \BibitemShut  [1]{\csname bibitem#1\endcsname}%
\let\auto@bib@innerbib\@empty
\bibitem [{\citenamefont {Wu}\ \emph {et~al.}(2020)\citenamefont {Wu},
  \citenamefont {Zhao}, \citenamefont {Yu}, \citenamefont {Chen}, \citenamefont
  {Wang}, \citenamefont {Song}, \citenamefont {Hu}, \citenamefont {Tao},
  \citenamefont {Tian}, \citenamefont {Pei} \emph {et~al.}}]{wu2020new}%
  \BibitemOpen
  \bibfield  {author} {\bibinfo {author} {\bibfnamefont {F.}~\bibnamefont
  {Wu}}, \bibinfo {author} {\bibfnamefont {S.}~\bibnamefont {Zhao}}, \bibinfo
  {author} {\bibfnamefont {B.}~\bibnamefont {Yu}}, \bibinfo {author}
  {\bibfnamefont {Y.-M.}\ \bibnamefont {Chen}}, \bibinfo {author}
  {\bibfnamefont {W.}~\bibnamefont {Wang}}, \bibinfo {author} {\bibfnamefont
  {Z.-G.}\ \bibnamefont {Song}}, \bibinfo {author} {\bibfnamefont
  {Y.}~\bibnamefont {Hu}}, \bibinfo {author} {\bibfnamefont {Z.-W.}\
  \bibnamefont {Tao}}, \bibinfo {author} {\bibfnamefont {J.-H.}\ \bibnamefont
  {Tian}}, \bibinfo {author} {\bibfnamefont {Y.-Y.}\ \bibnamefont {Pei}}, \emph
  {et~al.},\ }\bibfield  {title} {\bibinfo {title} {A new coronavirus
  associated with human respiratory disease in {C}hina},\ }\href@noop {}
  {\bibfield  {journal} {\bibinfo  {journal} {Nature}\ }\textbf {\bibinfo
  {volume} {579}},\ \bibinfo {pages} {265} (\bibinfo {year}
  {2020})}\BibitemShut {NoStop}%
\bibitem [{\citenamefont {Ciotti}\ \emph {et~al.}(2020)\citenamefont {Ciotti},
  \citenamefont {Ciccozzi}, \citenamefont {Terrinoni}, \citenamefont {Jiang},
  \citenamefont {Wang},\ and\ \citenamefont {Bernardini}}]{ciotti2020covid}%
  \BibitemOpen
  \bibfield  {author} {\bibinfo {author} {\bibfnamefont {M.}~\bibnamefont
  {Ciotti}}, \bibinfo {author} {\bibfnamefont {M.}~\bibnamefont {Ciccozzi}},
  \bibinfo {author} {\bibfnamefont {A.}~\bibnamefont {Terrinoni}}, \bibinfo
  {author} {\bibfnamefont {W.-C.}\ \bibnamefont {Jiang}}, \bibinfo {author}
  {\bibfnamefont {C.-B.}\ \bibnamefont {Wang}},\ and\ \bibinfo {author}
  {\bibfnamefont {S.}~\bibnamefont {Bernardini}},\ }\bibfield  {title}
  {\bibinfo {title} {The {COVID}-19 pandemic},\ }\href@noop {} {\bibfield
  {journal} {\bibinfo  {journal} {Critical reviews in clinical laboratory
  sciences}\ }\textbf {\bibinfo {volume} {57}},\ \bibinfo {pages} {365}
  (\bibinfo {year} {2020})}\BibitemShut {NoStop}%
\bibitem [{\citenamefont {Sanders}\ \emph {et~al.}(2020)\citenamefont
  {Sanders}, \citenamefont {Monogue}, \citenamefont {Jodlowski},\ and\
  \citenamefont {Cutrell}}]{sanders2020pharmacologic}%
  \BibitemOpen
  \bibfield  {author} {\bibinfo {author} {\bibfnamefont {J.~M.}\ \bibnamefont
  {Sanders}}, \bibinfo {author} {\bibfnamefont {M.~L.}\ \bibnamefont
  {Monogue}}, \bibinfo {author} {\bibfnamefont {T.~Z.}\ \bibnamefont
  {Jodlowski}},\ and\ \bibinfo {author} {\bibfnamefont {J.~B.}\ \bibnamefont
  {Cutrell}},\ }\bibfield  {title} {\bibinfo {title} {{Pharmacologic Treatments
  for Coronavirus Disease 2019 (COVID-19): A Review}},\ }\href
  {https://doi.org/10.1001/jama.2020.6019} {\bibfield  {journal} {\bibinfo
  {journal} {JAMA}\ }\textbf {\bibinfo {volume} {323}},\ \bibinfo {pages}
  {1824} (\bibinfo {year} {2020})}\BibitemShut {NoStop}%
\bibitem [{\citenamefont {Flaxman}\ \emph {et~al.}(2020)\citenamefont
  {Flaxman}, \citenamefont {Mishra}, \citenamefont {Gandy}, \citenamefont
  {Unwin}, \citenamefont {Coupland}, \citenamefont {Mellan}, \citenamefont
  {Zhu}, \citenamefont {Berah}, \citenamefont {Eaton}, \citenamefont
  {Perez~Guzman} \emph {et~al.}}]{flaxman2020report}%
  \BibitemOpen
  \bibfield  {author} {\bibinfo {author} {\bibfnamefont {S.}~\bibnamefont
  {Flaxman}}, \bibinfo {author} {\bibfnamefont {S.}~\bibnamefont {Mishra}},
  \bibinfo {author} {\bibfnamefont {A.}~\bibnamefont {Gandy}}, \bibinfo
  {author} {\bibfnamefont {H.}~\bibnamefont {Unwin}}, \bibinfo {author}
  {\bibfnamefont {H.}~\bibnamefont {Coupland}}, \bibinfo {author}
  {\bibfnamefont {T.}~\bibnamefont {Mellan}}, \bibinfo {author} {\bibfnamefont
  {H.}~\bibnamefont {Zhu}}, \bibinfo {author} {\bibfnamefont {T.}~\bibnamefont
  {Berah}}, \bibinfo {author} {\bibfnamefont {J.}~\bibnamefont {Eaton}},
  \bibinfo {author} {\bibfnamefont {P.}~\bibnamefont {Perez~Guzman}}, \emph
  {et~al.},\ }\href@noop {} {\emph {\bibinfo {title} {Report 13: Estimating the
  number of infections and the impact of non-pharmaceutical interventions on
  {COVID-19} in 11 European countries}}},\ \bibinfo {type} {Tech. Rep.}\
  (\bibinfo {year} {2020})\BibitemShut {NoStop}%
\bibitem [{\citenamefont {Perra}(2021)}]{perra2021non}%
  \BibitemOpen
  \bibfield  {author} {\bibinfo {author} {\bibfnamefont {N.}~\bibnamefont
  {Perra}},\ }\bibfield  {title} {\bibinfo {title} {Non-pharmaceutical
  interventions during the {COVID}-19 pandemic: A review},\ }\href
  {https://doi.org/https://doi.org/10.1016/j.physrep.2021.02.001} {\bibfield
  {journal} {\bibinfo  {journal} {Physics Reports}\ }\textbf {\bibinfo {volume}
  {913}},\ \bibinfo {pages} {1} (\bibinfo {year} {2021})}\BibitemShut {NoStop}%
\bibitem [{\citenamefont {Ndwandwe}\ and\ \citenamefont
  {Wiysonge}(2021)}]{ndwandwe2021covid}%
  \BibitemOpen
  \bibfield  {author} {\bibinfo {author} {\bibfnamefont {D.}~\bibnamefont
  {Ndwandwe}}\ and\ \bibinfo {author} {\bibfnamefont {C.~S.}\ \bibnamefont
  {Wiysonge}},\ }\bibfield  {title} {\bibinfo {title} {{COVID}-19 vaccines},\
  }\href@noop {} {\bibfield  {journal} {\bibinfo  {journal} {Current opinion in
  immunology}\ }\textbf {\bibinfo {volume} {71}},\ \bibinfo {pages} {111}
  (\bibinfo {year} {2021})}\BibitemShut {NoStop}%
\bibitem [{\citenamefont {Meehan}\ \emph {et~al.}(2020)\citenamefont {Meehan},
  \citenamefont {Rojas}, \citenamefont {Adekunle}, \citenamefont {Adegboye},
  \citenamefont {Caldwell}, \citenamefont {Turek}, \citenamefont {Williams},
  \citenamefont {Marais}, \citenamefont {Trauer},\ and\ \citenamefont
  {McBryde}}]{meehan2020modelling}%
  \BibitemOpen
  \bibfield  {author} {\bibinfo {author} {\bibfnamefont {M.~T.}\ \bibnamefont
  {Meehan}}, \bibinfo {author} {\bibfnamefont {D.~P.}\ \bibnamefont {Rojas}},
  \bibinfo {author} {\bibfnamefont {A.~I.}\ \bibnamefont {Adekunle}}, \bibinfo
  {author} {\bibfnamefont {O.~A.}\ \bibnamefont {Adegboye}}, \bibinfo {author}
  {\bibfnamefont {J.~M.}\ \bibnamefont {Caldwell}}, \bibinfo {author}
  {\bibfnamefont {E.}~\bibnamefont {Turek}}, \bibinfo {author} {\bibfnamefont
  {B.~M.}\ \bibnamefont {Williams}}, \bibinfo {author} {\bibfnamefont {B.~J.}\
  \bibnamefont {Marais}}, \bibinfo {author} {\bibfnamefont {J.~M.}\
  \bibnamefont {Trauer}},\ and\ \bibinfo {author} {\bibfnamefont {E.~S.}\
  \bibnamefont {McBryde}},\ }\bibfield  {title} {\bibinfo {title} {Modelling
  insights into the {COVID}-19 pandemic},\ }\href
  {https://doi.org/https://doi.org/10.1016/j.prrv.2020.06.014} {\bibfield
  {journal} {\bibinfo  {journal} {Paediatric Respiratory Reviews}\ }\textbf
  {\bibinfo {volume} {35}},\ \bibinfo {pages} {64} (\bibinfo {year}
  {2020})}\BibitemShut {NoStop}%
\bibitem [{\citenamefont {Vespignani}\ \emph {et~al.}(2020)\citenamefont
  {Vespignani}, \citenamefont {Tian}, \citenamefont {Dye}, \citenamefont
  {Lloyd-Smith}, \citenamefont {Eggo}, \citenamefont {Shrestha}, \citenamefont
  {Scarpino}, \citenamefont {Gutierrez}, \citenamefont {Kraemer}, \citenamefont
  {Wu} \emph {et~al.}}]{vespignani2020modelling}%
  \BibitemOpen
  \bibfield  {author} {\bibinfo {author} {\bibfnamefont {A.}~\bibnamefont
  {Vespignani}}, \bibinfo {author} {\bibfnamefont {H.}~\bibnamefont {Tian}},
  \bibinfo {author} {\bibfnamefont {C.}~\bibnamefont {Dye}}, \bibinfo {author}
  {\bibfnamefont {J.~O.}\ \bibnamefont {Lloyd-Smith}}, \bibinfo {author}
  {\bibfnamefont {R.~M.}\ \bibnamefont {Eggo}}, \bibinfo {author}
  {\bibfnamefont {M.}~\bibnamefont {Shrestha}}, \bibinfo {author}
  {\bibfnamefont {S.~V.}\ \bibnamefont {Scarpino}}, \bibinfo {author}
  {\bibfnamefont {B.}~\bibnamefont {Gutierrez}}, \bibinfo {author}
  {\bibfnamefont {M.~U.}\ \bibnamefont {Kraemer}}, \bibinfo {author}
  {\bibfnamefont {J.}~\bibnamefont {Wu}}, \emph {et~al.},\ }\bibfield  {title}
  {\bibinfo {title} {Modelling covid-19},\ }\href@noop {} {\bibfield  {journal}
  {\bibinfo  {journal} {Nature Reviews Physics}\ }\textbf {\bibinfo {volume}
  {2}},\ \bibinfo {pages} {279} (\bibinfo {year} {2020})}\BibitemShut {NoStop}%
\bibitem [{\citenamefont {Vaughan}(2021)}]{vaughan2021omicron}%
  \BibitemOpen
  \bibfield  {author} {\bibinfo {author} {\bibfnamefont {A.}~\bibnamefont
  {Vaughan}},\ }\bibfield  {title} {\bibinfo {title} {Omicron emerges},\ }\href
  {https://doi.org/https://doi.org/10.1016/S0262-4079(21)02140-0} {\bibfield
  {journal} {\bibinfo  {journal} {New Scientist}\ }\textbf {\bibinfo {volume}
  {252}},\ \bibinfo {pages} {7} (\bibinfo {year} {2021})}\BibitemShut {NoStop}%
\bibitem [{\citenamefont {Eker}(2020)}]{eker2020validity}%
  \BibitemOpen
  \bibfield  {author} {\bibinfo {author} {\bibfnamefont {S.}~\bibnamefont
  {Eker}},\ }\bibfield  {title} {\bibinfo {title} {Validity and usefulness of
  {COVID}-19 models},\ }\href@noop {} {\bibfield  {journal} {\bibinfo
  {journal} {Humanities and Social Sciences Communications}\ }\textbf {\bibinfo
  {volume} {7}},\ \bibinfo {pages} {1} (\bibinfo {year} {2020})}\BibitemShut
  {NoStop}%
\bibitem [{\citenamefont {Chadeau-Hyam}\ \emph {et~al.}(2022)\citenamefont
  {Chadeau-Hyam}, \citenamefont {Tang}, \citenamefont {Eales}, \citenamefont
  {Bodinier}, \citenamefont {Wang}, \citenamefont {Jonnerby}, \citenamefont
  {Whitaker}, \citenamefont {Elliott}, \citenamefont {Haw}, \citenamefont
  {Walters} \emph {et~al.}}]{chadeau2022omicron}%
  \BibitemOpen
  \bibfield  {author} {\bibinfo {author} {\bibfnamefont {M.}~\bibnamefont
  {Chadeau-Hyam}}, \bibinfo {author} {\bibfnamefont {D.}~\bibnamefont {Tang}},
  \bibinfo {author} {\bibfnamefont {O.}~\bibnamefont {Eales}}, \bibinfo
  {author} {\bibfnamefont {B.}~\bibnamefont {Bodinier}}, \bibinfo {author}
  {\bibfnamefont {H.}~\bibnamefont {Wang}}, \bibinfo {author} {\bibfnamefont
  {J.}~\bibnamefont {Jonnerby}}, \bibinfo {author} {\bibfnamefont
  {M.}~\bibnamefont {Whitaker}}, \bibinfo {author} {\bibfnamefont
  {J.}~\bibnamefont {Elliott}}, \bibinfo {author} {\bibfnamefont
  {D.}~\bibnamefont {Haw}}, \bibinfo {author} {\bibfnamefont {C.~E.}\
  \bibnamefont {Walters}}, \emph {et~al.},\ }\bibfield  {title} {\bibinfo
  {title} {Omicron {SARS}-{C}o{V}-2 epidemic in {E}ngland during {F}ebruary
  2022: A series of cross-sectional community surveys},\ }\href
  {https://doi.org/https://doi.org/10.1016/j.lanepe.2022.100462} {\bibfield
  {journal} {\bibinfo  {journal} {The Lancet Regional Health - Europe}\
  }\textbf {\bibinfo {volume} {21}},\ \bibinfo {pages} {100462} (\bibinfo
  {year} {2022})}\BibitemShut {NoStop}%
\bibitem [{\citenamefont {Sigal}\ \emph {et~al.}(2022)\citenamefont {Sigal},
  \citenamefont {Milo},\ and\ \citenamefont {Jassat}}]{sigal2022estimating}%
  \BibitemOpen
  \bibfield  {author} {\bibinfo {author} {\bibfnamefont {A.}~\bibnamefont
  {Sigal}}, \bibinfo {author} {\bibfnamefont {R.}~\bibnamefont {Milo}},\ and\
  \bibinfo {author} {\bibfnamefont {W.}~\bibnamefont {Jassat}},\ }\bibfield
  {title} {\bibinfo {title} {Estimating disease severity of {O}micron and
  {D}elta {SARS}-{C}o{V}-2 infections},\ }\href@noop {} {\bibfield  {journal}
  {\bibinfo  {journal} {Nature Reviews Immunology}\ }\textbf {\bibinfo {volume}
  {22}},\ \bibinfo {pages} {267} (\bibinfo {year} {2022})}\BibitemShut
  {NoStop}%
\bibitem [{\citenamefont {Verity}\ \emph {et~al.}(2020)\citenamefont {Verity},
  \citenamefont {Okell}, \citenamefont {Dorigatti}, \citenamefont {Winskill},
  \citenamefont {Whittaker}, \citenamefont {Imai}, \citenamefont
  {Cuomo-Dannenburg}, \citenamefont {Thompson}, \citenamefont {Walker},
  \citenamefont {Fu} \emph {et~al.}}]{verity2020estimates}%
  \BibitemOpen
  \bibfield  {author} {\bibinfo {author} {\bibfnamefont {R.}~\bibnamefont
  {Verity}}, \bibinfo {author} {\bibfnamefont {L.~C.}\ \bibnamefont {Okell}},
  \bibinfo {author} {\bibfnamefont {I.}~\bibnamefont {Dorigatti}}, \bibinfo
  {author} {\bibfnamefont {P.}~\bibnamefont {Winskill}}, \bibinfo {author}
  {\bibfnamefont {C.}~\bibnamefont {Whittaker}}, \bibinfo {author}
  {\bibfnamefont {N.}~\bibnamefont {Imai}}, \bibinfo {author} {\bibfnamefont
  {G.}~\bibnamefont {Cuomo-Dannenburg}}, \bibinfo {author} {\bibfnamefont
  {H.}~\bibnamefont {Thompson}}, \bibinfo {author} {\bibfnamefont {P.~G.}\
  \bibnamefont {Walker}}, \bibinfo {author} {\bibfnamefont {H.}~\bibnamefont
  {Fu}}, \emph {et~al.},\ }\bibfield  {title} {\bibinfo {title} {Estimates of
  the severity of coronavirus disease 2019: a model-based analysis},\ }\href
  {https://doi.org/https://doi.org/10.1016/S1473-3099(20)30243-7} {\bibfield
  {journal} {\bibinfo  {journal} {The Lancet Infectious Diseases}\ }\textbf
  {\bibinfo {volume} {20}},\ \bibinfo {pages} {669} (\bibinfo {year}
  {2020})}\BibitemShut {NoStop}%
\bibitem [{\citenamefont {Rodrigues}\ and\ \citenamefont
  {Helene}(2020)}]{rodrigues2020monte}%
  \BibitemOpen
  \bibfield  {author} {\bibinfo {author} {\bibfnamefont {T.}~\bibnamefont
  {Rodrigues}}\ and\ \bibinfo {author} {\bibfnamefont {O.}~\bibnamefont
  {Helene}},\ }\bibfield  {title} {\bibinfo {title} {Monte carlo approach to
  model {COVID}-19 deaths and infections using gompertz functions},\
  }\href@noop {} {\bibfield  {journal} {\bibinfo  {journal} {Physical Review
  Research}\ }\textbf {\bibinfo {volume} {2}},\ \bibinfo {pages} {043381}
  (\bibinfo {year} {2020})}\BibitemShut {NoStop}%
\bibitem [{\citenamefont {Chen}\ \emph {et~al.}(2022)\citenamefont {Chen},
  \citenamefont {Yan}, \citenamefont {Sun}, \citenamefont {Zheng},
  \citenamefont {Sun}, \citenamefont {Zhou}, \citenamefont {Deng},
  \citenamefont {Zhuang}, \citenamefont {Cai}, \citenamefont {Zhang},
  \citenamefont {Ajelli},\ and\ \citenamefont {Yu}}]{chen2022estimation}%
  \BibitemOpen
  \bibfield  {author} {\bibinfo {author} {\bibfnamefont {X.}~\bibnamefont
  {Chen}}, \bibinfo {author} {\bibfnamefont {X.}~\bibnamefont {Yan}}, \bibinfo
  {author} {\bibfnamefont {K.}~\bibnamefont {Sun}}, \bibinfo {author}
  {\bibfnamefont {N.}~\bibnamefont {Zheng}}, \bibinfo {author} {\bibfnamefont
  {R.}~\bibnamefont {Sun}}, \bibinfo {author} {\bibfnamefont {J.}~\bibnamefont
  {Zhou}}, \bibinfo {author} {\bibfnamefont {X.}~\bibnamefont {Deng}}, \bibinfo
  {author} {\bibfnamefont {T.}~\bibnamefont {Zhuang}}, \bibinfo {author}
  {\bibfnamefont {J.}~\bibnamefont {Cai}}, \bibinfo {author} {\bibfnamefont
  {J.}~\bibnamefont {Zhang}}, \bibinfo {author} {\bibfnamefont
  {M.}~\bibnamefont {Ajelli}},\ and\ \bibinfo {author} {\bibfnamefont
  {H.}~\bibnamefont {Yu}},\ }\bibfield  {title} {\bibinfo {title} {Estimation
  of disease burden and clinical severity of {COVID}-19 caused by {O}micron
  {BA}.2 in {S}hanghai, {F}ebruary-{J}une 2022},\ }\href
  {https://doi.org/10.1080/22221751.2022.2128435} {\bibfield  {journal}
  {\bibinfo  {journal} {Emerging Microbes \& Infections}\ }\textbf {\bibinfo
  {volume} {11}},\ \bibinfo {pages} {2800} (\bibinfo {year} {2022})},\ \bibinfo
  {note} {pMID: 36205530}\BibitemShut {NoStop}%
\bibitem [{\citenamefont {Erikstrup}\ \emph {et~al.}(2022)\citenamefont
  {Erikstrup}, \citenamefont {Laksafoss}, \citenamefont {Gladov}, \citenamefont
  {Kaspersen}, \citenamefont {Mikkelsen}, \citenamefont {Hindhede},
  \citenamefont {Boldsen}, \citenamefont {J{\o}rgensen}, \citenamefont
  {Ethelberg}, \citenamefont {Holm} \emph
  {et~al.}}]{erikstrup2022seroprevalence}%
  \BibitemOpen
  \bibfield  {author} {\bibinfo {author} {\bibfnamefont {C.}~\bibnamefont
  {Erikstrup}}, \bibinfo {author} {\bibfnamefont {A.~D.}\ \bibnamefont
  {Laksafoss}}, \bibinfo {author} {\bibfnamefont {J.}~\bibnamefont {Gladov}},
  \bibinfo {author} {\bibfnamefont {K.~A.}\ \bibnamefont {Kaspersen}}, \bibinfo
  {author} {\bibfnamefont {S.}~\bibnamefont {Mikkelsen}}, \bibinfo {author}
  {\bibfnamefont {L.}~\bibnamefont {Hindhede}}, \bibinfo {author}
  {\bibfnamefont {J.~K.}\ \bibnamefont {Boldsen}}, \bibinfo {author}
  {\bibfnamefont {S.~W.}\ \bibnamefont {J{\o}rgensen}}, \bibinfo {author}
  {\bibfnamefont {S.}~\bibnamefont {Ethelberg}}, \bibinfo {author}
  {\bibfnamefont {D.~K.}\ \bibnamefont {Holm}}, \emph {et~al.},\ }\bibfield
  {title} {\bibinfo {title} {Seroprevalence and infection fatality rate of the
  {SARS}-{C}o{V}-2 {O}micron variant in {D}enmark: A nationwide
  serosurveillance study},\ }\href@noop {} {\bibfield  {journal} {\bibinfo
  {journal} {The Lancet Regional Health-Europe}\ }\textbf {\bibinfo {volume}
  {21}},\ \bibinfo {pages} {100479} (\bibinfo {year} {2022})}\BibitemShut
  {NoStop}%
\bibitem [{\citenamefont {Helene}\ \emph {et~al.}(2016)\citenamefont {Helene},
  \citenamefont {Mariano},\ and\ \citenamefont
  {Guimaraes-Filho}}]{helene2016useful}%
  \BibitemOpen
  \bibfield  {author} {\bibinfo {author} {\bibfnamefont {O.}~\bibnamefont
  {Helene}}, \bibinfo {author} {\bibfnamefont {L.}~\bibnamefont {Mariano}},\
  and\ \bibinfo {author} {\bibfnamefont {Z.}~\bibnamefont {Guimaraes-Filho}},\
  }\bibfield  {title} {\bibinfo {title} {Useful and little-known applications
  of the least square method and some consequences of covariances},\
  }\href@noop {} {\bibfield  {journal} {\bibinfo  {journal} {Nuclear
  Instruments and Methods in Physics Research Section A: Accelerators,
  Spectrometers, Detectors and Associated Equipment}\ }\textbf {\bibinfo
  {volume} {833}},\ \bibinfo {pages} {82} (\bibinfo {year} {2016})}\BibitemShut
  {NoStop}%
\bibitem [{\citenamefont {Cuadros}\ \emph {et~al.}(2022)\citenamefont
  {Cuadros}, \citenamefont {Moreno}, \citenamefont {Musuka}, \citenamefont
  {Miller}, \citenamefont {Coule},\ and\ \citenamefont
  {MacKinnon}}]{cuadros2022association}%
  \BibitemOpen
  \bibfield  {author} {\bibinfo {author} {\bibfnamefont {D.~F.}\ \bibnamefont
  {Cuadros}}, \bibinfo {author} {\bibfnamefont {C.~M.}\ \bibnamefont {Moreno}},
  \bibinfo {author} {\bibfnamefont {G.}~\bibnamefont {Musuka}}, \bibinfo
  {author} {\bibfnamefont {F.~D.}\ \bibnamefont {Miller}}, \bibinfo {author}
  {\bibfnamefont {P.}~\bibnamefont {Coule}},\ and\ \bibinfo {author}
  {\bibfnamefont {N.~J.}\ \bibnamefont {MacKinnon}},\ }\bibfield  {title}
  {\bibinfo {title} {Association between vaccination coverage disparity and the
  dynamics of the {COVID}-19 {D}elta and {O}micron waves in the {US}},\
  }\href@noop {} {\bibfield  {journal} {\bibinfo  {journal} {Frontiers in
  Medicine}\ }\textbf {\bibinfo {volume} {9}} (\bibinfo {year}
  {2022})}\BibitemShut {NoStop}%
\bibitem [{\citenamefont {Barletta}(2022)}]{barletta2022influence}%
  \BibitemOpen
  \bibfield  {author} {\bibinfo {author} {\bibfnamefont {W.~A.}\ \bibnamefont
  {Barletta}},\ }\bibfield  {title} {\bibinfo {title} {The {I}nfluence of
  {SARS}-{C}o{V}-2 {V}ariants on {N}ational {C}ase-{F}atality {R}ates:
  {C}orrelation and {V}alidation {S}tudy},\ }\href@noop {} {\bibfield
  {journal} {\bibinfo  {journal} {JMIRx Med}\ }\textbf {\bibinfo {volume}
  {3}},\ \bibinfo {pages} {e32935} (\bibinfo {year} {2022})}\BibitemShut
  {NoStop}%
\bibitem [{\citenamefont {Emani}\ \emph {et~al.}(2022)\citenamefont {Emani},
  \citenamefont {Pallipuram}, \citenamefont {Goswami}, \citenamefont {Maddula},
  \citenamefont {Reddy}, \citenamefont {Nakka}, \citenamefont {Panga},
  \citenamefont {Reddy}, \citenamefont {Reddy}, \citenamefont {Nandanoor} \emph
  {et~al.}}]{emani2022increasing}%
  \BibitemOpen
  \bibfield  {author} {\bibinfo {author} {\bibfnamefont {V.~R.}\ \bibnamefont
  {Emani}}, \bibinfo {author} {\bibfnamefont {V.~K.}\ \bibnamefont
  {Pallipuram}}, \bibinfo {author} {\bibfnamefont {K.~K.}\ \bibnamefont
  {Goswami}}, \bibinfo {author} {\bibfnamefont {K.~R.}\ \bibnamefont
  {Maddula}}, \bibinfo {author} {\bibfnamefont {R.}~\bibnamefont {Reddy}},
  \bibinfo {author} {\bibfnamefont {A.~S.}\ \bibnamefont {Nakka}}, \bibinfo
  {author} {\bibfnamefont {S.}~\bibnamefont {Panga}}, \bibinfo {author}
  {\bibfnamefont {N.~K.}\ \bibnamefont {Reddy}}, \bibinfo {author}
  {\bibfnamefont {N.~K.}\ \bibnamefont {Reddy}}, \bibinfo {author}
  {\bibfnamefont {D.}~\bibnamefont {Nandanoor}}, \emph {et~al.},\ }\bibfield
  {title} {\bibinfo {title} {Increasing sars-cov2 cases, hospitalizations, and
  deaths among the vaccinated populations during the omicron (b. 1.1. 529)
  variant surge in uk},\ }\href@noop {} {\bibfield  {journal} {\bibinfo
  {journal} {medRxiv}\ ,\ \bibinfo {pages} {2022}} (\bibinfo {year}
  {2022})}\BibitemShut {NoStop}%
\bibitem [{\citenamefont {Wang}\ \emph {et~al.}(2023)\citenamefont {Wang},
  \citenamefont {Liu}, \citenamefont {Zhang}, \citenamefont {Huang},
  \citenamefont {Zhao}, \citenamefont {Lu},\ and\ \citenamefont
  {Cui}}]{wang2022differences}%
  \BibitemOpen
  \bibfield  {author} {\bibinfo {author} {\bibfnamefont {C.}~\bibnamefont
  {Wang}}, \bibinfo {author} {\bibfnamefont {B.}~\bibnamefont {Liu}}, \bibinfo
  {author} {\bibfnamefont {S.}~\bibnamefont {Zhang}}, \bibinfo {author}
  {\bibfnamefont {N.}~\bibnamefont {Huang}}, \bibinfo {author} {\bibfnamefont
  {T.}~\bibnamefont {Zhao}}, \bibinfo {author} {\bibfnamefont {Q.-B.}\
  \bibnamefont {Lu}},\ and\ \bibinfo {author} {\bibfnamefont {F.}~\bibnamefont
  {Cui}},\ }\bibfield  {title} {\bibinfo {title} {Differences in incidence and
  fatality of {COVID}-19 by {SARS}-{C}o{V}-2 {O}micron variant versus {D}elta
  variant in relation to vaccine coverage: A world-wide review},\ }\href
  {https://doi.org/https://doi.org/10.1002/jmv.28118} {\bibfield  {journal}
  {\bibinfo  {journal} {Journal of Medical Virology}\ }\textbf {\bibinfo
  {volume} {95}},\ \bibinfo {pages} {e28118} (\bibinfo {year}
  {2023})}\BibitemShut {NoStop}%
\bibitem [{\citenamefont {Kim}\ \emph {et~al.}(2023)\citenamefont {Kim},
  \citenamefont {Cho}, \citenamefont {Song}, \citenamefont {Rahmati},
  \citenamefont {Koyanagi}, \citenamefont {Lee}, \citenamefont {Yon},
  \citenamefont {Il~Shin},\ and\ \citenamefont {Smith}}]{kim2023case}%
  \BibitemOpen
  \bibfield  {author} {\bibinfo {author} {\bibfnamefont {K.}~\bibnamefont
  {Kim}}, \bibinfo {author} {\bibfnamefont {K.}~\bibnamefont {Cho}}, \bibinfo
  {author} {\bibfnamefont {J.}~\bibnamefont {Song}}, \bibinfo {author}
  {\bibfnamefont {M.}~\bibnamefont {Rahmati}}, \bibinfo {author} {\bibfnamefont
  {A.}~\bibnamefont {Koyanagi}}, \bibinfo {author} {\bibfnamefont {S.~W.}\
  \bibnamefont {Lee}}, \bibinfo {author} {\bibfnamefont {D.~K.}\ \bibnamefont
  {Yon}}, \bibinfo {author} {\bibfnamefont {J.}~\bibnamefont {Il~Shin}},\ and\
  \bibinfo {author} {\bibfnamefont {L.}~\bibnamefont {Smith}},\ }\bibfield
  {title} {\bibinfo {title} {The case fatality rate of {COVID}-19 during the
  {D}elta and the {O}micron epidemic phase: A meta-analysis},\ }\href
  {https://doi.org/https://doi.org/10.1002/jmv.28522} {\bibfield  {journal}
  {\bibinfo  {journal} {Journal of Medical Virology}\ }\textbf {\bibinfo
  {volume} {95}},\ \bibinfo {pages} {e28522} (\bibinfo {year}
  {2023})}\BibitemShut {NoStop}%
\end{thebibliography}%

\begin{figure*}

\includegraphics[scale=0.67]{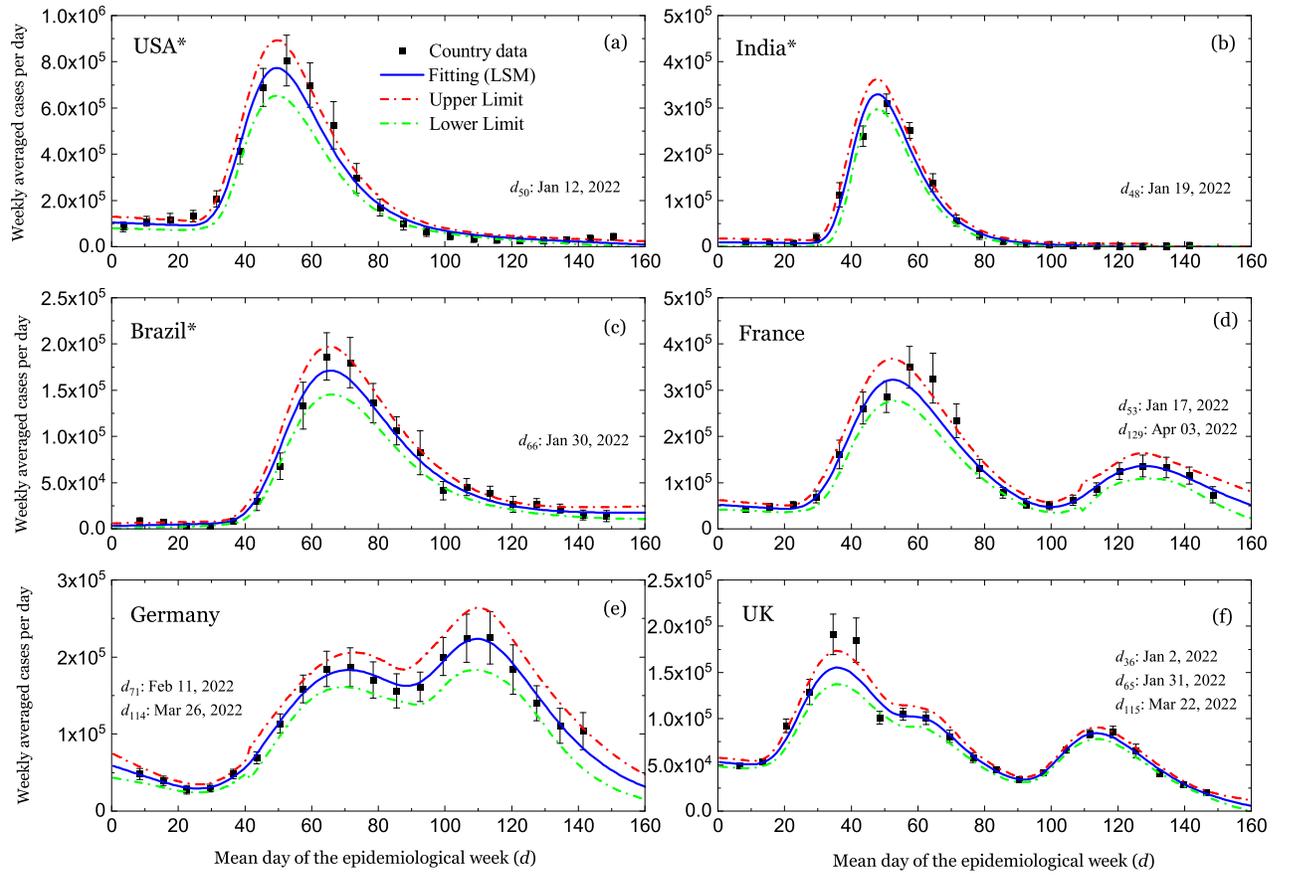}
\caption{\label{fig:Fig1} Weekly-averaged cases per day for USA (a), India (b), Brazil (c), France (d), Germany (e) and UK (f) (data points with error bars) and the respective fitting using the LSM (solid blue lines) with its upper and lower 95\% confidence intervals given by the dashed-dotted red and green lines, respectively. (*) For USA, India and Brazil additional constant errors of, respectively, 1\%, 2\% and 1\% of its corresponding peak heights were added quadratically with the standard errors in order to achieve a successful fitting. Also shown the dates of each peak of the Gompertz functions.}
\end{figure*}

\begin{figure*}
\includegraphics[scale=0.67]{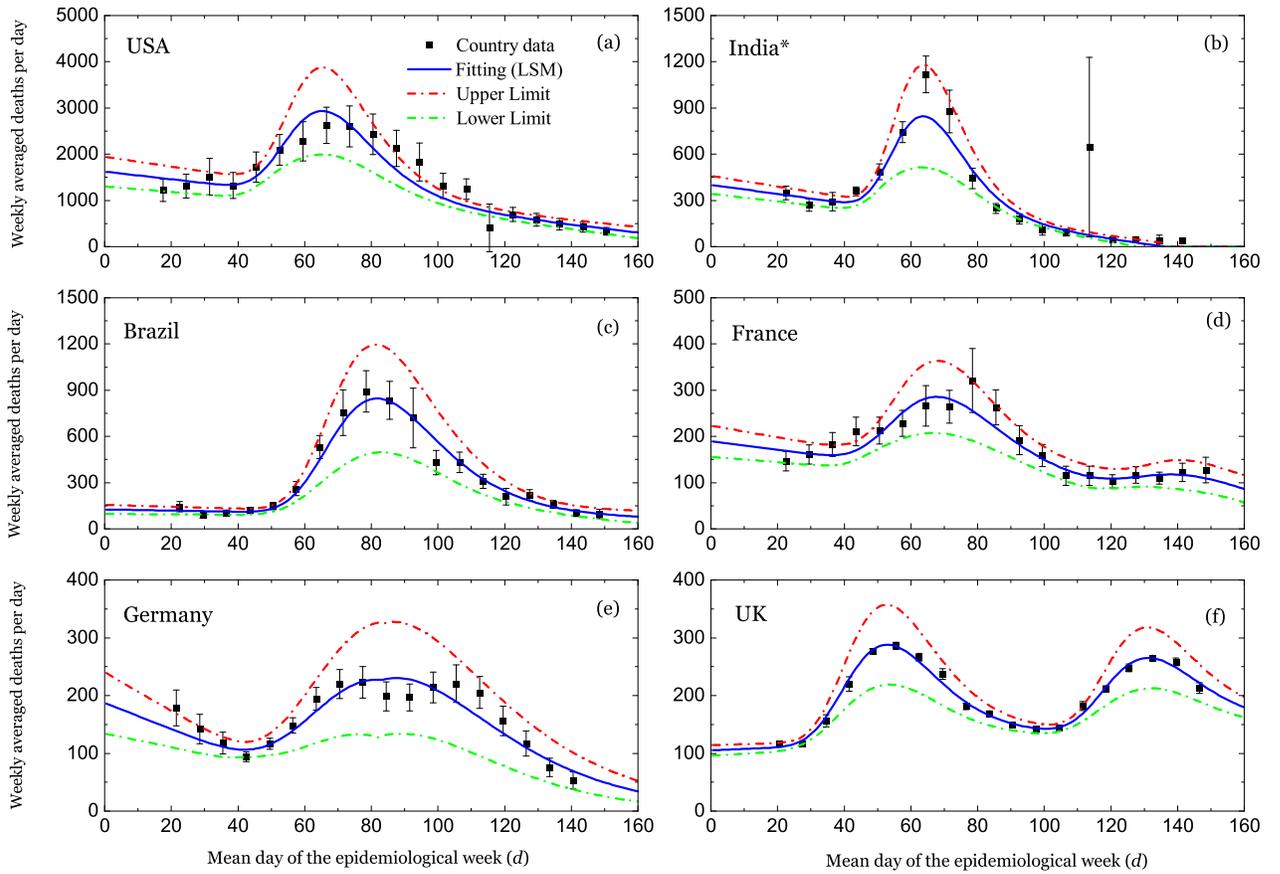}
\caption{\label{fig:Fig2}Weekly-averaged deaths per day for USA (a), India (b), Brazil (c), France (d), Germany (e) and UK (f) (data points with error bars) and the respective fitting using the LSM (solid blue lines) with its upper and lower 95\% confidence intervals given by the dashed-dotted red and green lines, respectively. (*) For India, an additional error of 1\% of its corresponding peak height was added quadratically with the standard errors in order to achieve a successful fitting.}
\end{figure*}

\begin{figure*}
\includegraphics[scale=0.67]{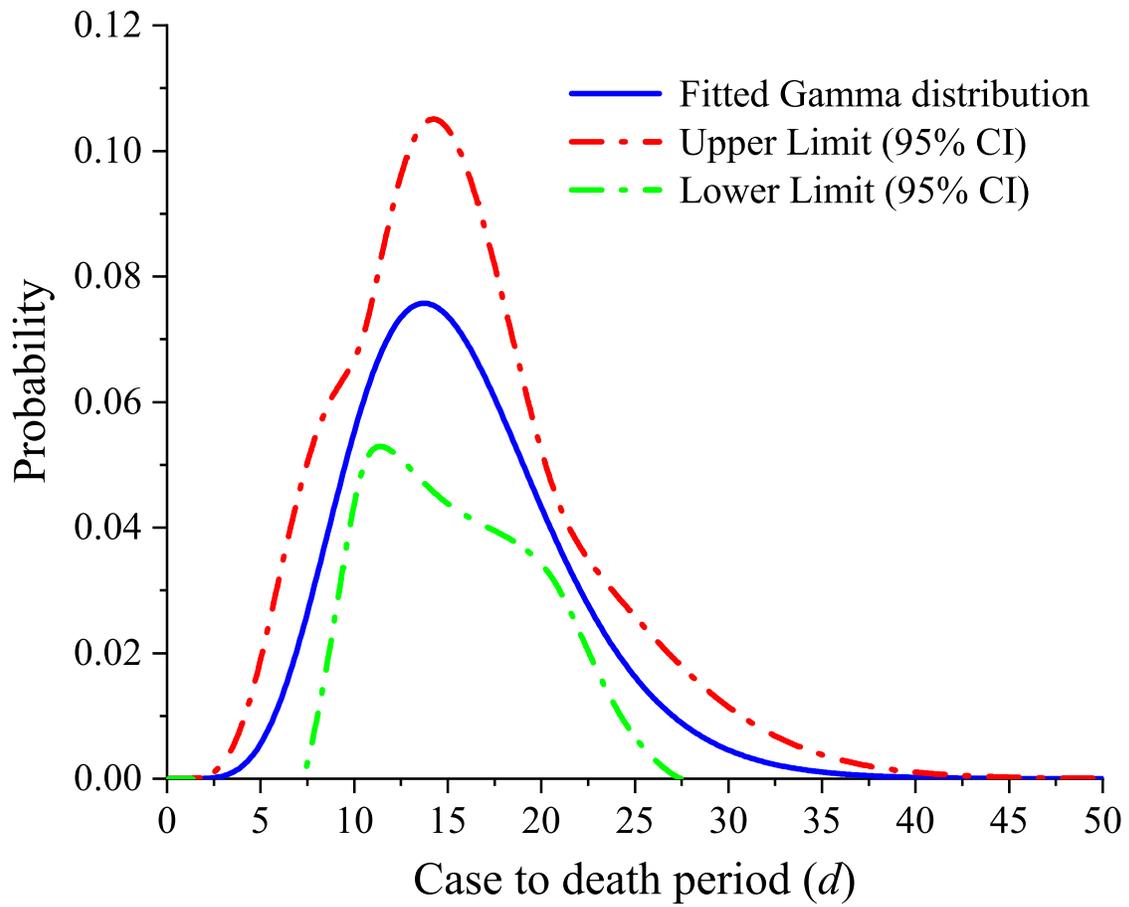}
\caption{\label{fig:Fig3}Fitted gamma function for the case to death period (solid blue line) and its upper (dashed-dotted red) and lower (dashed-dotted green) 95\% confidence intervals.}
\end{figure*}

\begin{figure*}
\includegraphics[scale=0.67]{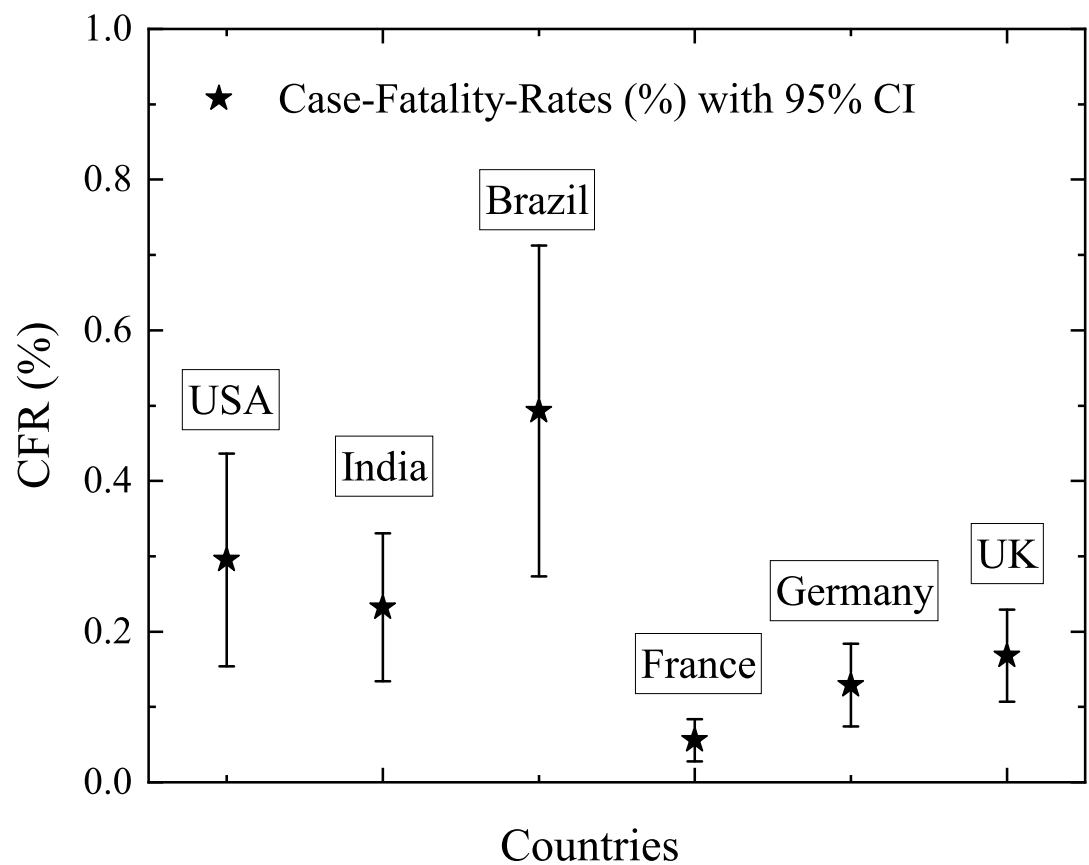}
\caption{\label{fig:Fig4}Case-Fatality-Rates (\%) and respective 95\% confidence intervals for the USA, India, Brazil, France, Germany and United Kingdom.}
\end{figure*}

\begin{table*}[b]
\caption{Best-fit parameters of the Gompertz functions and linear backgrounds for the weekly-averaged daily confirmed cases for each country according to Monte Carlo simulations. Also shown the lowest $\chi^{2}$, the number of degrees of freedom ($N.D.F$), the probability of exceeding the $\chi^{2}$ (Prob) and the date of the first Omicron case for each country.}
\label{table1}

\resizebox{\textwidth}{!}{\begin{tabular}{cccccccc}
    \hline
\multirow{2}{*}{Parameters}&  \multicolumn{6}{c}{Countries}\\
& USA $^a$&	India $^b$ &	Brazil $^a$&	France &	Germany &	UK\\

 \hline
   \multirow{3}{*}{$N_{c}(\times 10^{6})$ } & 21.8 &	7.70&	6.74 &11.9	&11.8&	3.95 \\
     &  &  & & 6.26	&6.44	&0.917\\
      &  &  & & & & 2.20 \\
       \multirow{3}{*}{$\lambda (d^{-1})$ } & 0.0925 &	0.115&	0.0673 &0.0696	&0.0421&	0.0848 \\
     &  &  & & 0.0574	&0.0636	&0.119\\
      &  &  & & & & 0.0843 \\
      \multirow{3}{*}{$t_{0} (d)$ } & 49.4 &	48.8&	66.9 &52.7	&70.0&	34.4 \\
     &  &  & & 128	&113	&64.6\\
      &  &  & & & & 114 \\
     $C_{c}(\times 10^{4})$ &10.8	&0.827&	0.345&	5.23&	5.83&	5.13 \\
      $S_{c}(d^{-1})(\times 100)$ &-6,07&	-0,639&	0,793&	-4,72&	-13,3&	-2,95
 \\
       $\chi^{2}$ &19.1&	7.53&	12.3&	9.62&	2.84&	12.3\\
       $N.D.F.$ &17&15	&16	&13	&12&	10\\
      Prob (\%) &32.1&	94.1&	72.5&	72.5&	99.7&	26.3\\
      First Omicron case & Nov 22, 2021 & Dec 02, 2021 & Nov 25, 2021 & Nov 25, 2021& Dec 02, 2021 & Nov 27, 2021\\
      \hline
  \end{tabular}}\\
  \footnotesize{$^a$ An additional 1\% error was included in the weekly-averaged daily cases data in order to achieve a successful fitting.} \\
  \footnotesize{$^b$  An additional 2\% error was included in the weekly-averaged daily cases data in order to achieve a successful fitting.}\\
\end{table*}

\begin{table*}[b]
\caption{Best-fit parameters of the Gompertz functions and linear backgrounds for the weekly-averaged daily cases and deaths for each country according to the LSM. Also shown the case-fatality-rates (CFR), the fitted parameters of the gamma function ($\alpha $ and $\beta$ and its corresponding $\mu$ and CV), the global $\chi^{2}$, the number of degrees of freedom  ($N.D.F.$) and the probability of exceeding the $\chi^{2}$ (Prob).}
\label{table2}
  \resizebox{\textwidth}{!}{\begin{tabular}{ccccccc}
    \hline
\multirow{2}{*}{Parameters}&  \multicolumn{6}{c}{Countries}\\
& USA $^a$ &	India $^b$ &	Brazil $^a$ &	France &	Germany &	UK\\

 \hline
   \multirow{3}{*}{$N_{c}(\times 10^{6})$ } &22.0$\pm$1.6 &	7.68$\pm$0.39 &	6.70$\pm$0.46 &	11.9$\pm$1.2 &	12.56$\pm$0.96 &3.86$\pm$0.31\\
     &  &  & & 6.26$\pm$0.78 &	5.85$\pm$0.98 &	0.88$\pm$0.16\\
      &  &  & & & & 2.35$\pm$0.26\\
       \multirow{3}{*}{$\lambda (d^{-1})$ } & 0.0863$\pm$0.0051 &	0.1147$\pm$0.0059 &	0.0660$\pm$0.0036 &	0.0675$\pm$0.0054 &	0.0396$\pm$0.0023 &	0.0803$\pm$0.0046\\
     &  &  & & 0.057$\pm$0.010 &	0.066$\pm$0.011 & 0.113$\pm$0.017\\
      &  &  & & & & 0.0784$\pm$0.0056\\
      \multirow{3}{*}{$t_{0} (d)$ } & 49.70$\pm$0.74 &	47.85$\pm$0.48 &	65.53$\pm$0.84 &	52.8$\pm$1.1 &	71.3$\pm$1.8 &	36.26$\pm$0.66 \\
     &  &  & & 128.8$\pm$2.1 &	113.6$\pm$2.2 &	65.2$\pm$1.6\\
      &  &  & & & & 114.57$\pm$0.68 \\
     $C_{c}(\times 10^{4})$ &10.5$\pm$1.3 &	0.93$\pm$0.43 &	0.30$\pm$0.15 &	5.23$\pm$0.51 &	5.93$\pm$0.79 &	5.31$\pm$0.24 \\
      $S_{c}(d^{-1})(\times 100)$ &-6.0$\pm$1.0 &	-0.71$\pm$0.45 &	0.85$\pm$0.27 &	-4.8$\pm$2.0 &	-14.3$\pm$4.1 &	-3.29$\pm$0.32 \\
      \multirow{2}{*}{$N_{d} (\times 10^{4})$ }&6.5$\pm$1.5 &	1.78$\pm$0.38 &	3.30$\pm$0.71 &	0.68$\pm$0.17 &	1.62$\pm$0.34 &	0.63$\pm$0.12 \\
      & & & &0.28$\pm$0.16 & & 0.48$\pm$0.10\\
      $C_{d}(\times 100)$ &16.2$\pm$1.6 &	3.98$\pm$0.29 &	1.27$\pm$0.15 &	1.90$\pm$0.17 &	1.86$\pm$0.27 &	1.05$\pm$0.05 \\
      $S_{d}(d^{-1})$ &-8.2$\pm$1.2	& -2.95$\pm$0.26 &	-0.37$\pm$0.20 &	-0.91$\pm$0.25 &	-2.26$\pm$0.62 &	0.244$\pm$0.059\\
      \multirow{2}{*}{CFR $(\%)$ }&0.295$\pm$0.071 &	0.232$\pm$0.049	& 0.49$\pm$0.11 &	0.057$\pm$0.014	& 0.129$\pm$0.027 &	0.164$\pm$0.031 \\
       & & & &0.043$\pm$0.025	& &	0.205$\pm$0.035\\
       $\alpha$ &\multicolumn{6}{c}{8.0$\pm$3.1}\\
       $\beta (d^{-1})$ &\multicolumn{6}{c}{0.51$\pm$0.21}\\
       $\mu (d)$ &\multicolumn{6}{c}{15.71$\pm$0.55}\\
       CV &\multicolumn{6}{c}{0.354$\pm$0.070}\\
       $\chi^{2}$ &\multicolumn{6}{c}{176.5}\\
       $N.D.F.$ &\multicolumn{6}{c}{174}\\
      Prob (\%) &\multicolumn{6}{c}{43.4}\\
      \hline 
  \end{tabular}}
  \footnotesize{$^a$ An additional 1\% error was included in the weekly-averaged daily cases data in order to achieve a successful fitting.} \\
  \footnotesize{$^b$  An additional 2\% error was included in the weekly-averaged daily cases and deaths data in order to achieve a successful fitting.}\\
\end{table*}

\end{document}